\documentclass[11pt,notitlepage,nofootinbib]{revtex4-1}

\usepackage{epsf}
\usepackage{latexsym}
\usepackage{amssymb}
\usepackage{amsmath}
\usepackage{bm}
\usepackage[dvips]{graphicx}
\usepackage{array}
\usepackage{xcolor}

\date{\today}
\def\d3{^{(3)}\nabla}

\begin{document}

\title{Magnetic field back reaction on the matter power spectrum}

\author{Kerstin E. Kunze}

\email{kkunze@usal.es}

\affiliation{Departamento de F\'\i sica Fundamental, Universidad de Salamanca,
 Plaza de la Merced s/n, 37008 Salamanca, Spain}

\begin{abstract}
At lowest order comoving  magnetic fields which are frozen-into the expanding cosmic fluid do not evolve in time.
At next-to-leading order the induction equation is sourced by the interaction term between the baryon velocity and the magnetic field amplitude which leads to 
a non-trivial evolution of the comoving magnetic field.
Moreover, it induces non-trivial cross correlations between the adiabatic curvature mode and the magnetic mode.
This cross correlation together with the evolution of the induced matter perturbation leads to interesting effects on the total matter power spectrum at small
scales.
\end{abstract}

\maketitle

\section{Introduction}
\label{s0}
\setcounter{equation}{0}

Direct  observational evidence for the existence of cosmological magnetic fields is hard to come by.
Cosmological data indicate that the universe is well described by the standard $\Lambda$CDM model which
 rules out the presence of strong magnetic fields on large scales.
Assuming that these magnetic fields were generated in the early universe long before decoupling there are different possibilities to check for their imprints on physical observables. Such magnetic fields can be generated, e.g., during inflation or during a phase transition.
These mechanisms allow to generate stochastic magnetic fields which generally are assumed to be gaussian.
Cosmological magnetic fields originating in the early universe could play an important role later on at the time of 
galaxy formation and serve as initial seed field for the presently observed magnetic fields in spiral galaxies (for reviews, e.g., \cite{Durrer:2013pga,rev4}).

Primordial magnetic fields have an effect on the temperature anisotropies and polarization angular power spectra.
They contribute to the total energy density perturbation as well as the anisotropic stress and the Lorentz term contributes to the evolution equation of the baryon velocity. Other effects include Faraday rotation of the linearly polarized CMB signal (cf., e.g., \cite{Giovannini:2004aw, Kosowsky:2004zh,
kb, Giovannini:2008aa,pfp,sl1, kk11,Pogosian:2011qv}).

Magnetic fields are embedded in the cosmic fluid leading to effects on the fluid variables as well as the magnetic field. 
At lowest order  the magnetic field can be treated as frozen-in into the cosmic fluid so that its energy density is redshifted in the expanding universe the  same way  as radiation. This allows to rescale the amplitude of the magnetic field and define a comoving magnetic field whose spectrum does not evolve in time.
Taking into account the back reaction of the peculiar velocity in the induction equation leads to a non trivial evolution of the magnetic field.
This implies a cross correlation with the baryon density perturbation.
The effect on the temperature and polarization anisotropies of the cosmic microwave background (CMB)  in this case have been studied in \cite{kk13}.
Here the focus is on the effect on the matter power spectrum.

\section{Evolution of the  magnetic field}
\label{s2}
\setcounter{equation}{0}

In a cosmological background with metric $ds^2=a^2(\eta)(-d\eta^2+d{\mathbf x}^2)$ with scale factor $a(\eta)$ and $\eta$ conformal time,
the evolution of the magnetic field in real space is determined by the induction equation in the form,
\begin{eqnarray}
\partial_{\eta}(a^2{\bf B})({\bf x}, \eta)={\bf \nabla}\times\left[{\bf V}_b({\bf x}, \eta)\times(a^2{\bf B})({\bf x}, \eta)\right]
\label{e1}
\end{eqnarray}
where ${\bf V}_b({\bf x}, \eta)$ is the baryon velocity.
Equation (\ref{e1}) yields to 
\begin{eqnarray}
\partial_{\eta}\left(a^2{\bf B}\right)=\left(a^2{\bf B}\cdot{\bf \nabla}\right){\bf V}_b-\left({\bf V}_b\cdot{\bf \nabla}\right)\left(a^2{\bf B}\right)-a^2{\bf B}\left({\bf \nabla}\cdot{\bf V}_b\right)
\label{e2}
\end{eqnarray}
Expanding  ${\bf B}({\bf x}, \eta)$ and ${\bf V}_b({\bf x}, \eta)$ in terms of scalar and vector spherical harmonics, as follows,
\begin{eqnarray}
a^2B_i({\bf x}, \eta)=a_0^2\sum_{\bf k}B_i({\bf k},\eta) Q^{(0)}({\bf k},{\bf x})
\end{eqnarray}
and
\begin{eqnarray}
V_{b,i}({\bf x}, \eta)=\sum_{m=0,\pm 1}\sum_{\bf k} V^{(m)}({\bf k},\eta)Q_i^{(m)}({\bf k},{\bf x}),
\end{eqnarray}
where $i=1,2,3$.
The scalar harmonics are $Q^{(0)}({\bf k},{\bf x})=e^{i{\bf k}\cdot{\bf x}}$ together with the  definition $Q_i^{(0)}=-k^{-1}{\bf \nabla}_iQ^{(0)}=-i\frac{k_i}{k}e^{i{\bf k}\cdot{\bf x}}$. The vector harmonics are conveniently represented in the helicity basis as
$Q_i^{(\pm 1)}({\bf k},{\bf x})=\left(\hat{\bf e}_{\bf k}^{(\pm 1)}\right)_ie^{i{\bf k}\cdot{\bf x}}$ where $\hat{\bf e}_{\bf k}^{(\pm 1)}=-\frac{i}{\sqrt{2}}(\hat{\bf e}_{\bf k}^{(1)}\pm i \hat{\bf e}_{\bf k}^{(2)})$ and with $\hat{\bf e}_{\bf k}^{(3)}=\frac{\bf k}{k}$ \cite{hw,kk13}.  
Using this in equation (\ref{e2}) leads to 
\begin{eqnarray}
\dot{B}_i({\bf k},\eta)&=&\sum_{\bf q}\left[\frac{q_ik_j-{\bf k}\cdot{\bf q}\delta_{ij}}{q}B_j({\bf k}-{\bf q},\eta)V^{(0)}_b({\bf q},\eta)
\right.\nonumber\\
&&
\left.
+i\left[k_j(\hat{\bf e}_{\bf q}^{(+1)})_i-(\hat{\bf e}_{\bf q}^{(+1)})_mk_m\delta_{ij}\right]B_j({\bf k}-{\bf q},\eta)V^{(+1)}_b({\bf q},\eta)
\right.\nonumber\\
&& 
\left.
+i\left[k_j(\hat{\bf e}_{\bf q}^{(-1)})_i-(\hat{\bf e}_{\bf q}^{(-1)})_mk_m\delta_{ij}\right]B_j({\bf k}-{\bf q},\eta)V^{(-1)}_b({\bf q},\eta)
\right]
\end{eqnarray}
where a dot denotes the derivative w.r.t. to conformal time.
Whereas the scalar mode of the baryon velocity is sourced by the adiabatic, primordial curvature mode as well as by the 
 magnetic mode, the vector mode only receives a significant contribution from the 
magnetic mode. However, the latter is proportional to the anisotropic stress part which is quadratic in the magnetic field amplitude.
Moreover, already at first order the correction to the linear matter power spectrum will enter as a 6-point function in $B_i({\bf k},\eta)$ which is proportional to the magnetic field power spectrum cubed. This follows from the same arguments as the contribution of the magnetic vector mode to the
angular power spectrum of the CMB anisotropies \cite{kk13}. Therefore, in the following the contribution of the magnetic vector mode will be neglected.

For the scalar mode the baryon velocity follows the evolution
(e.g. \cite{kk11}, \cite{sl1}) 
\begin{eqnarray}
\dot{V}_{\rm b}^{(0)}=(3c_s^2-1){\cal H}V_{\rm b}^{(0)}+k(\Psi-3c_s^2\Phi)+kc_s^2\Delta_{\rm b}+R\tau_c^{-1}(V_{\gamma}^{(0)}-V_{\rm b})^{(0)}+\frac{R}{4}kL^{(0)},
\label{vb}
\end{eqnarray}
where, in terms of the magnetic energy density $\Delta_B$ is defined by
\begin{eqnarray}
\Delta_B(\vec{k},\tau)=\frac{1}{2\rho_{\gamma 0}}\sum_{\vec{q}}B_i(\vec{q},\tau)B^i(\vec{k}-\vec{q},\tau).
\end{eqnarray}
and the anisotropic stress $\pi_B^{(0)}$ is defined by
\begin{eqnarray}
\pi_B^{(0)}(\vec{k},\tau)&=&\frac{3}{2\rho_{\gamma0}}\left[\sum_{\vec{q}}\frac{3}{k^2}B_i(\vec{k}-\vec{q},\tau)q^iB_j(\vec{q},\tau)\left(k^j-q^j\right)-\sum_{\vec{q}}B_m(\vec{k}-\vec{q},\tau)B^m(\vec{q},\tau)\right],
\end{eqnarray}
the Lorentz term $L^{(0)}=\Delta_B-\frac{2}{3}\pi_B^{(0)}$ is due to the Lorentz force $\vec{J}\times\vec{B}$ and $R\equiv\frac{4}{3}\frac{\rho_{\gamma}}{\rho_{\rm b}}$. Furthermore, $c_s^2=\frac{\partial\bar{p}}{\partial\bar{\rho}}$ is the adiabatic sound speed and $\tau_c^{-1}$ is the mean free path of photons between scatterings given in terms of the number density of free electrons $n_{\rm e}$ and the Thomson cross section $\sigma_{\rm T}$, $\tau_c^{-1}=an_{\rm e}\sigma_{\rm T}$. Moreover, $\Phi$ and $\Psi$ denote the gauge invariant Bardeen potentials and ${\cal H}=\frac{\dot{a}}{a}$. As can be seen for example in the baryon velocity equation, the magnetic field does not enter linearly, but rather quadratically. 
This is also the case for the remaining perturbation equations (e.g., \cite{kk11,sl1,kk12}).

During the matter dominated era $V_b^{(0)}=-k^{-1}\dot{\Delta}_b$ thus the evolution of the comoving magnetic field amplitude due to the scalar mode
of the velocity perturbation is given by
\begin{eqnarray}
\dot{B}_i({\bf k},\eta)=-\sum_{\bf q}\frac{q_iq_j-{\bf k}\cdot{\bf q}\delta_{ij}}{q^2}B_j({\bf k}-{\bf q},\eta)\dot{\Delta}_b({\bf q},\eta).
\label{dotB}
\end{eqnarray}
It is interesting to note that equation (\ref{dotB}) can be solved formally using 
an iterative expansion scheme similar to the one applied  in cosmological perturbation theory (cf., e.g., \cite{ggrw}). 
Writing the magnetic field as  a sum of the form $\sum_n a(\eta)^n p_n({\bf k})$ where
$a(\eta)$ is the scale factor and $p_n({\bf k})$ an expansion coefficient only depending on the wave number then 
\begin{eqnarray}
B_i({\bf k},\eta)=\sum_{n=0}^{\infty}\left(\frac{\eta}{\eta_0}\right)^{2n}F_{i,n}({\bf k})
\label{Bexp}
\end{eqnarray}
using that during matter domination the scale factor evolves as $(\eta/\eta_0)^2$ and $\eta_0$ is chosen to be the present epoch.
In general the  total baryon density perturbation  is assumed to be of the form
\begin{eqnarray}
\Delta_b({\bf q},\eta)=\sum_{m_1=0}^{\infty}\left(\frac{\eta}{\eta_0}\right)^{2m_1}\Delta_{b,m_1}^{(ad)}({\bf q},\eta_0)+
\sum_{m_2=0}^{\infty}\left(\frac{\eta}{\eta_0}\right)^{2m_2}\Delta_{b,m_2}^{(B)}({\bf q},\eta_0)
\end{eqnarray}
where $(ad)$ indicates the contribution from the adiabatic, primordial curvature perturbation and $(B)$ the contributions form the (compensated) magnetic mode. Taking into account  only the growing mode of the linear density perturbation yields to contributions  only from  
the terms for $m_1=1$ and $m_2=1$. It can be shown that the growing mode of the linear, magnetic perturbation evolves with the same growth factor as the adiabatic mode, namely, proportional to $\eta^2$ during the matter dominated era (e.g., \cite{sesu,kk14}).
Using these expansions in equation (\ref{dotB}) the coefficients $F_{i,n}$ in the expansion of the magnetic field (\ref{Bexp}) are found to be
\begin{eqnarray}
\delta_{i,i_n}F_{i,n}({\bf k})&=&\frac{(-1)^n}{n!}\sum_{{\bf q}_1}...\sum_{{\bf q}_n}\left[\prod_{\beta=1}^n
\frac{q_{\beta,i_{\beta}}q_{\beta,i_{\beta-1}}-({\bf k}-{\bf Q}+\sum_{\alpha=1}^{\beta}{\bf q}_{\alpha})\cdot{\bf q}_{\beta}\delta_{i_{\beta},i_{\beta-1}}}
{q_{\beta}^2}
\right.
\nonumber\\
&&\left.
\cdot\left[\Delta_b^{(ad)}({\bf q}_{\beta},\eta_0)+\Delta_b^{(B)}({\bf q}_{\beta},\eta_0)\right]\right]F_{i_0,0}({\bf k}-{\bf Q}),
\end{eqnarray}
where ${\bf Q}\equiv\sum_{\alpha=1}^n{\bf q}_{\alpha}$ and the indices $i_{\mu}=1,2,3$.
Since at lowest order the comoving magnetic field is constant in time $F_{i,0}({\bf k})$ is chosen to be the initial value at some initial time $\eta_i$.
Thus $F_{i,0}({\bf k})=B_i({\bf k},\eta_i)=B_i^{(0)}({\bf k})$. The initial time $\eta_i$ will be chosen to be at the beginning of the matter dominated epoch, at radiation-matter equality, $\eta_{eq}$. At this time higher order terms in the sum for $n\geq 1$ are suppressed by $\left(\frac{\eta_{eq}}{\eta_0}\right)^n$.

In the following the solution for the magnetic field will be used at first order in the iteration.

\section{Matter perturbations induced by an evolving comoving magnetic field}
\label{s2a}
\setcounter{equation}{0}

In the matter dominated universe on subhorizon scale the total matter perturbation is defined by 
\begin{eqnarray}
\Delta_m=\tilde{R}_c\Delta_c+\tilde{R}_b\Delta_b
\end{eqnarray}
where $\tilde{R}_b\equiv\frac{\rho_b}{\rho_b+\rho_c}$, $\tilde{R}_c\equiv\frac{\rho_c}{\rho_b+\rho_c}$, and $b$ and $c$ refer to baryons and cold dark
matter, respectively.
It evolves as (e.g., \cite{shawlew, kk14})
\begin{eqnarray}
\ddot{\Delta}_m+\frac{2}{\eta}\dot{\Delta}_m-\frac{6}{\eta^2}\Delta_m=-\frac{k^2}{3}\frac{\Omega_{\gamma,0}}{\Omega_{m,0}}\frac{\eta_0^2}{\eta^2}L({\mathbf k},\eta)
\label{Dm}
\end{eqnarray}
where $L({\mathbf k},\eta)$ is the Lorentz term related to the Lorentz force ${\mathbf L}({\mathbf x}, \eta)=\left[\left(\nabla\times{\mathbf B}\right)\times{\mathbf B}\right]({\mathbf x},\eta)$ and $\eta$ denotes conformal time. The index 0 refers to present time.
The growing mode solution of (\ref{Dm}) is given by
\begin{eqnarray}
\Delta_m({\mathbf k},\eta)&=&\alpha_{1,{\mathbf k}}\left(\frac{\eta}{\eta_0}\right)^2\nonumber\\
&-&\frac{(k\eta_0)^2}{15}\frac{\Omega_{\gamma,0}}{\Omega_{m,0}}\eta^2
\int_{\eta_i}^{\eta}d\hat{\eta}\hat{\eta}^{-3}
\left[1-\left(\frac{\hat{\eta}}{\eta}\right)^5\right]
L({\mathbf k},\hat{\eta})
\label{Dm-gen}
\end{eqnarray}
where $\alpha_{1,{\mathbf k}}$ is  constant and 
 $\eta_i$ is some initial time which is chosen to be the time of radiation-matter equality, $\eta_{eq}$.

 At lowest order of the iterative solution of the magnetic field the density perturbations of the adiabatic, primordial curvature mode and the magnetic mode are uncorrelated. At first order, however, there is a non vanishing correlation which is due to the presence of the amplitude of the baryon density perturbation in the first order solution of the magnetic field given explicitly, by $B_i({\mathbf k},\eta)=B^{(0)}_i({\mathbf k})+b^{(1)}_i({\mathbf k},\eta)$, where
\begin{eqnarray}
b^{(1)}_i({\mathbf k},\eta)=-\sum_{\mathbf q}\frac{q_ik_j-{\mathbf q}\cdot{\mathbf k}\delta_{ij}}{q^2}B^{(0)}_j({\mathbf k}-{\mathbf q})\left[\Delta_b^{(ad)}({\mathbf q},\eta_0)+\Delta_b^{(B)}({\mathbf q},\eta_0)\right]f(\eta)
\label{b1}
\end{eqnarray}
where $f(\eta)=\left(\frac{\eta}{\eta_0}\right)^2-\left(\frac{\eta_{eq}}{\eta_0}\right)^2$ using the evolution of the baryonic matter density perturbation at lowest order for the magnetic mode. This has the same growth factor as the adiabatic, primordial curvature mode (cf., e.g. \cite{kk14}).
Using the first oder solution of the magnetic field to calculate the Lorentz term is written as the lowest order,  time-independent part 
$L^{(0)}({\bf k})$ and the first order, time dependent part ${\cal L}^{(1)}({\bf k},\eta)$, such that 
\begin{eqnarray}
L({\mathbf k},\eta)=L^{(0)}({\mathbf k})+{\cal L}^{(1)}({\mathbf k},\eta).
\label{L}
\end{eqnarray}
Using this in the expressions for the matter density perturbations equation (\ref{Dm-gen})  leads to  cross correlations between the adiabatic and magnetic mode at times $\eta_a$ and $\eta_b$, respectively.
The resulting cross correlation functions between the adiabatic and magnetic matter density modes are determined 
by the cross correlation 2-point function between the time dependent part of the Lorentz term and the adiabatic mode,
written schematically as, 
\begin{eqnarray}
\langle\Delta_m^{(ad)*}({\bf k}',\eta_a)\Delta_m^{({\cal L}^{(1)})}({\bf k},\eta_b)\rangle=
{\cal F}[\eta_b,\langle\Delta_b^{(ad)*}({\bf k}',\eta_a){\cal L}^{(1)}({\bf k},\eta)\rangle]
\label{DeladDelb1}
\end{eqnarray}
where ${\cal F}[\eta_b,\langle\Delta_m^{(ad)*}({\bf k}',\eta_a){\cal L}({\bf k},\eta)\rangle]$
denotes the corresponding expression involving the integration over conformal time (cf. equation (\ref{Dm-gen})). The cross correlation between the adiabatic density perturbation and the Lorentz term at first order upto 
${\cal O}(\left(\frac{\rho_{B,0}}{\rho_{\gamma,0}}\right) )$ is given by
\begin{eqnarray}
\langle&\Delta_b^{(ad)*}&({\bf k}',\eta_a){\cal L}^{(1)}({\bf k},\eta)\rangle=-\frac{3}{2}\frac{f(\eta)}{\Gamma\left(\frac{n_B+3}{2}\right)}\left(\frac{\rho_{B,0}}{\rho_{\gamma,0}}\right)P^{(ad)}_m(k)\delta_{{\bf k}',{\bf k}}\left(\frac{k}{k_m}\right)^{n_B+3}
\nonumber\\
&\times&
\int_0^{\infty}dh h^2e^{-\left(\frac{k}{k_m}\right)^2h^2}
\int_{-1}^1dx\left[
h^{n_B}\left(-1-x^2+2hx^3\right)
\right.
\nonumber\\
&+&
\left.e^{-\left(\frac{k}{k_m}\right)^2(1-2hx)}\left(1-2hx+h^2\right)^{\frac{n_B-2}{2}}
h\left(-h-2x+5hx^2-2h^2x^3\right)\right]
\label{XDb}
\end{eqnarray}
where $x\equiv\frac{\bf k\cdot q}{kq}$ and  $h\equiv\frac{q}{k}$.

The total matter density perturbation is given by
\begin{eqnarray}
\Delta_m({\bf k},\eta)=\Delta_m^{(ad)}({\bf k},\eta)+\Delta_m^{(B^{(0)})}({\bf k},\eta)
+\Delta_m^{({\cal L}^{(1)})}({\bf k},\eta)
\label{Dmtotal}
\end{eqnarray}
where $\Delta_m^{(B^{(0)})}({\bf k},\eta)$  and $\Delta_m^{({\cal L}^{(1)})}({\bf k},\eta)$ are obtained when using
equation (\ref{L}) in (\ref{Dm-gen}) and separating the two contributions accordingly. 
The former  corresponds to the matter density perturbation by the magnetic mode associated with the frozen-in magnetic field which is assumed to be not correlated with the adabiatic, primordial curvature mode. The latter due to the evolution of the magnetic field results in a non vanishing cross correlation between the adiabatic and the magnetic mode
(cf equation (\ref{XDb})). The auto correlation function of the total matter power spectrum  at redshifts $z_a$ and 
$z_b$, respectively, is found to be 
\begin{eqnarray}
\langle\Delta_m^*({\bf k}',z_a)\Delta_m({\bf k},z_b)\rangle=\delta_{{\bf k}, {\bf k}'}D(z_a)D(z_b)\left[P_m^{(ad)}(k)
+P_m^{(B^{(0)})}(k)+P_m^{\langle\Delta_b^{(ad)}{\cal L}^{(1)}\rangle}(k,z_A,z_b)\right]
\label{delmdelm}
\end{eqnarray}
where $D(z)=1/(1+z)$ is the standard growth factor.  
The matter power spectrum of the adiabatic mode used in the numerical solutions is given by 
\cite{bbks,pu,hu,hw1,kk14}
\begin{eqnarray}
P^{(ad)}_m(k)=\frac{2\pi^2}{k^3}\frac{4}{25}A_s\left(\frac{k}{k_p}\right)^{n_s-1}T^2(k)
\left(\frac{k}{a_0H_0}\right)^4
\end{eqnarray}
with the transfer function
\begin{eqnarray}
T(k)=\frac{\ln(1+2.34q)}{2.34q}\left[1+3.89q+(16.1q)^2+(5.46q)^3+(6.71q)^4\right]^{-\frac{1}{4}},
\end{eqnarray}
with $q\equiv\frac{k}{\Omega_{m,0}h^2 {\rm Mpc}^{-1}}$
and the amplitude, spectral index and pivot wave number of the primordial curvature perturbation,  
$A_s$, $n_s$ and $k_p$, respectively.
The frozen-in magnetic field is assumed to be a nonhelical, gaussian random field with  its two-point function in Fourier space given by
(e.g., \cite{kk11})
\begin{eqnarray}
\langle B_i^*(\vec{k})B_j(\vec{q})\rangle=(2\pi)^3\delta({\vec{k}-\vec{q}})P_B(k)\left(\delta_{ij}-\frac{k_ik_j}{k^2}\right),
\end{eqnarray}
where the power spectrum, $P_B(k)$, is assumed to be a power law,
$P_B(k)=A_Bk^{n_B}$, with amplitude, $A_B$, and spectral index, $n_B$. Using the ensemble averaged
magnetic energy density, $\langle\rho_{B,0}\rangle$ at present,
applying a Gaussian window function and a smoothing scale $k_m$
the spectral function can be  expressed
as 
\begin{eqnarray}
P_{B,0}(k)=\frac{4\pi^2}{k_m^3}\frac{2^{(n_B+3)/2}}{\Gamma\left(\frac{n_B+3}{2}\right)}\left(\frac{k}{k_m}\right)^{n_B}\langle\rho_{B,0}\rangle.
\end{eqnarray}
Prior to decoupling in a process similar 
to the damping of density perturbations by photon diffusion, i.e. Silk damping, yields a 
maximal wave number $k_m$ determined by the Alfv\'en velocity and photon diffusion scale at decoupling
 \cite{jko,sb}         
\begin{eqnarray}
k_m=301.45\left(\frac{B_0}{\rm nG}\right)^{-1}{\rm Mpc}^{-1}
\end{eqnarray}
calculated \cite{kk21}  for the Planck 2018 best fit values of the six-parameter base $\Lambda$CDM model from Planck data alone \cite{Planck-2018}.

The matter power spectrum induced by the frozen-in magnetic field ${\bf B}^{(0)}({\bf k})$ is given by \cite{kk14}
\begin{eqnarray}
P_m^{(B^{(0)})}(k)=\frac{2\pi^2}{k^3}\left(\frac{k}{a_0H_0}\right)^4\frac{4}{225}\left(1+z_{eq}\right)^2
\left(\frac{\Omega_{\gamma,0}}{\Omega_{m,0}}\right)^2{\cal P}_{L^{(0)}}(k).
\end{eqnarray}
which is obtained from the zeroth order contribution $L^{(0)}(\vec{k})$ in the integral in
equation (\ref{Dm-gen}) .
The dimensionless power spectrum determining the two-point function of the Lorentz term 
$L^{(0)}({\bf k})$ is given by \cite{kk12}
\begin{eqnarray}
{\cal P}_{L^{(0)}}(k)&=&\frac{9}{\left[\Gamma\left(\frac{n_B+3}{2}\right)\right]^2}\left(\frac{\rho_{B,0}}{\rho_{\gamma,0}}\right)^2\left(\frac{k}{k_m}\right)^{2(n_B+3)}e^{-\left(\frac{k}{k_m}\right)^2}\nonumber\\
&\times&\int_0^{\infty}dh h^{n_B+2}e^{-2\left(\frac{k}{k_m}\right)^2h^2}\int_{-1}^1 dxe^{2\left(\frac{k}{k_m}\right)^2hx}
(1-2hx+h^2)^{\frac{n_B-2}{2}}\nonumber\\
&\times&\left[1+2h^2+(1-4h^2)x^2-4hx^3+4h^2x^4\right],
\label{pL}
\end{eqnarray}
and $x\equiv\frac{\bm{k}\cdot\bm{q}}{kq}$ and $h\equiv\frac{q}{k}$ where $\bm{q}$ is the wave number over which the 
resulting convolution integral is calculated. Moreover, using the ensemble averaged energy density of the magnetic field leads to $\frac{\rho_{B,0}}{\rho_{\gamma,0}}=9.545\times 10^{-8}\left(\frac{B_0}{\rm nG}\right)^2$ (cf. e.g. \cite{kk21}).

The cross correlation spectral function is found to be
\begin{eqnarray}
P_m^{\langle\Delta_b^{(ad)}{\cal L}^{(1)}\rangle}(k,z_a,z_b)&=&\frac{1}{5}\left(\frac{\Omega_{\gamma,0}}{\Omega_{m,0}}\right)\left(\frac{\rho_{B,0}}{\rho_{\gamma,0}}\right)\left(\frac{k}{a_0H_0}\right)^2P_m^{(ad)}(k)
{\cal I}_{\langle\Delta_b^{(ad)}{\cal L}^{(1)}\rangle}(k)
\nonumber\\
&\times&
\left[\ln\left(\frac{1+z_{eq}}{\sqrt{1+z_a}\sqrt{1+z_b}} \right)-\frac{7}{5}+\frac{5}{6}
\left[\left(\frac{1+z_{a}}{1+z_{eq}}\right)^{\frac{1}{2}}+\left(\frac{1+z_b}{1+z_{eq}}\right)^{\frac{1}{2}}\right]
\right.
\nonumber\\
&-&\left.\frac{2}{15}\left[\left(\frac{1+z_a}{1+z_{eq}}\right)^{\frac{5}{2}}+\left(\frac{1+z_b}{1+z_{eq}}\right)^{\frac{5}{2}}
\right]
\right]
\label{adL1}
\end{eqnarray}
where with the same definitions as above
\begin{eqnarray}
{\cal I}_{\langle\Delta_b^{(ad)}{\cal L}^{(1)}\rangle}(k)&=&-\frac{4}{3}+\frac{1}{\Gamma\left(\frac{n_B+3}{2}\right)}
\left(\frac{k}{k_m}\right)^{n_B+3}e^{-\left(\frac{k}{k_m}\right)^2}\int_0^{\infty}dh h^3
e^{-\left(\frac{k}{k_m}\right)^2h^2}
\int_{-1}^1 dx e^{2\left(\frac{k}{k_m}\right)^2hx}
\nonumber\\
&&\times
\left(1-2hx+h^2\right)^{\frac{n_B-2}{2}}
\left(-h-2x+5hx^2-2h^2x^3\right).
\end{eqnarray}

\begin{figure}[h!]
\centerline{\epsfxsize=3.75in\epsfbox{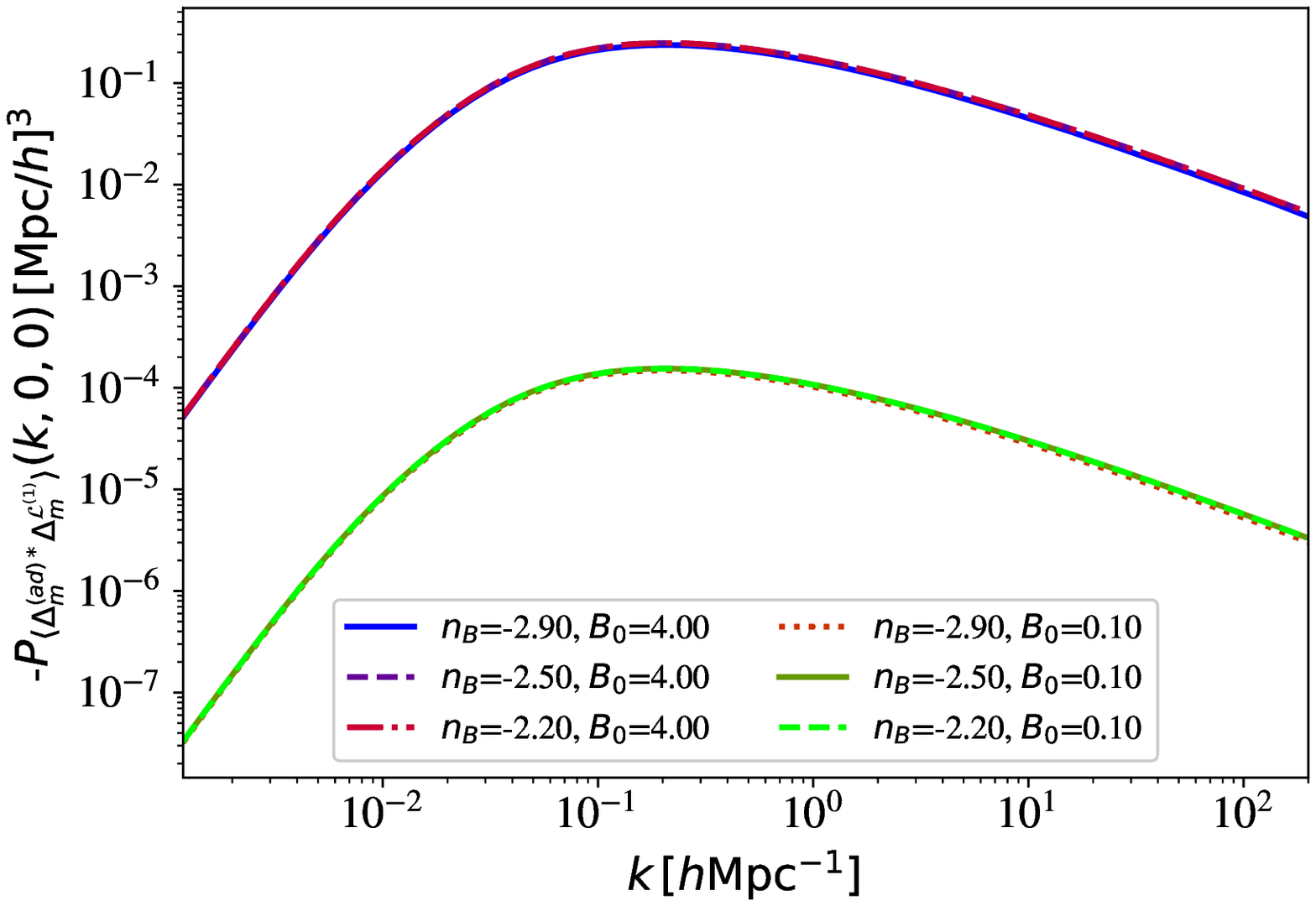}
\epsfxsize=3.75in\epsfbox{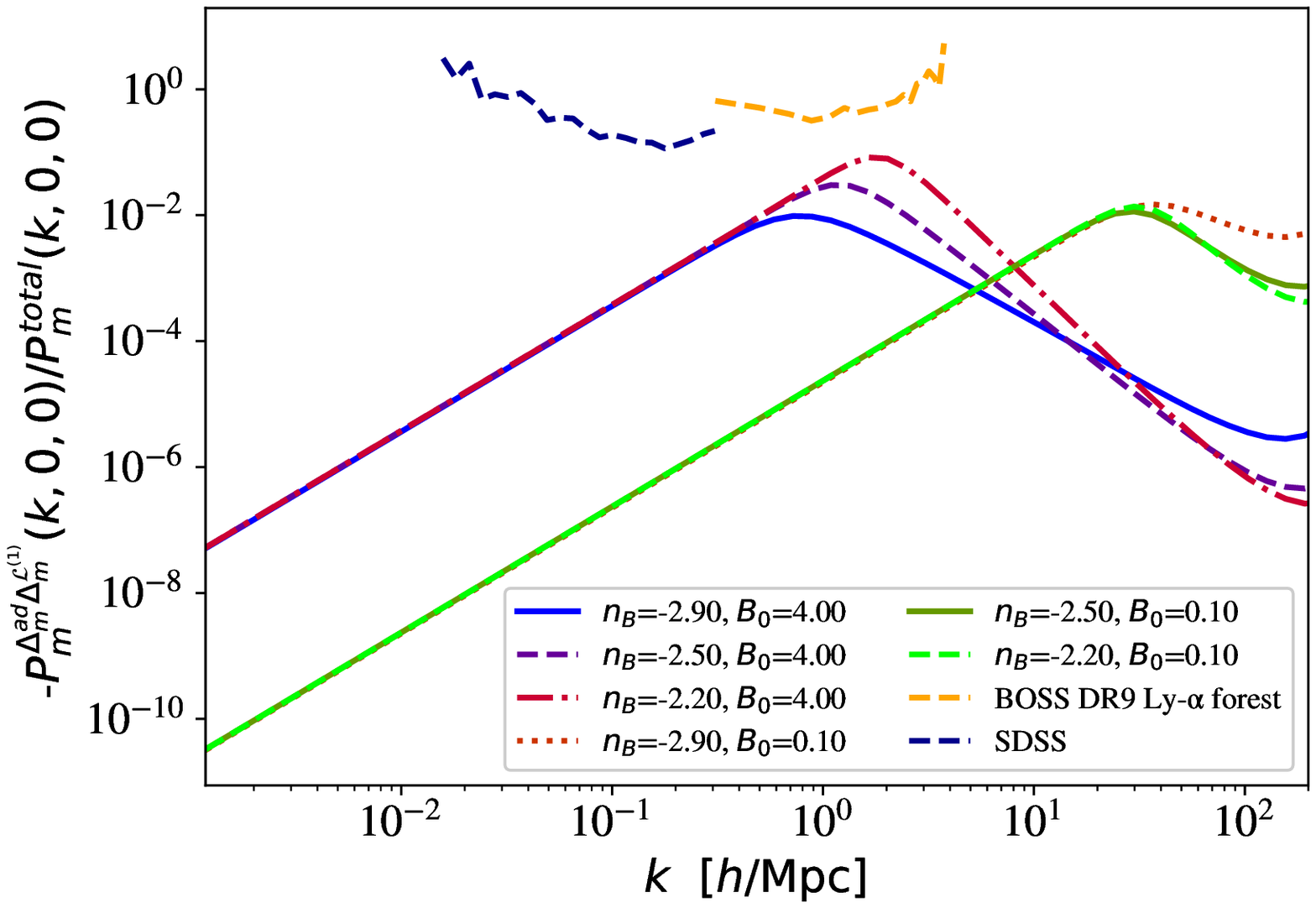}}
\centerline{\epsfxsize=3.75in\epsfbox{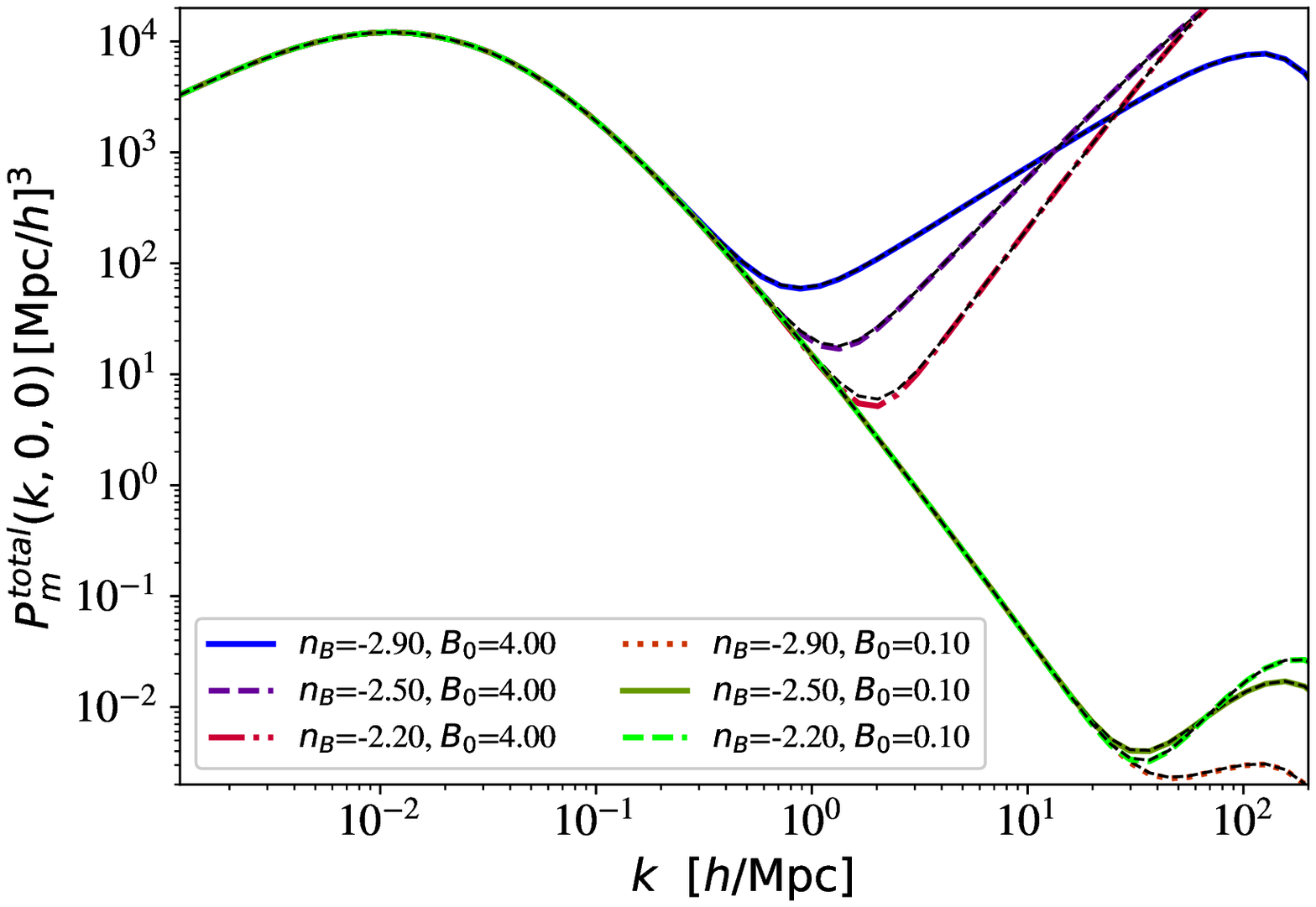}}
%\centerline{\epsfxsize=3.75in\epsfbox{1fig-PmadL1-eqTime_varBnBkhMpc.eps}
%\epsfxsize=3.75in\epsfbox{1fig-PmadL1rat-eqTime_varBnBkhMpc-totalwdat.eps}}
%\centerline{\epsfxsize=3.75in\epsfbox{1fig-Pmtotal-eqTime_varBnBkhMpc.eps}}
\caption{{\it Upper panel:} The equal time spectral function at present 
for magnetic field parameters ($B_0$ [nG], $n_B$)
determining the cross correlation function between the matter density perturbations due to the adiabatic mode and the evolving magnetic field mode   ({\it left}) and its ratio w.r.t. the total matter density autocorrelation function ({\it right}). In the latter the relative errors 
$\Delta P_m/P_m$ are included from BOSS DR9 Ly-$\alpha$ forrest data \cite{bossLyalpha} 
and SDSS data \cite{sdss}.
{\it Lower panel:} Equal time  spectral function at present determining the  total matter density autocorrelation function. The light black dashed lines show the autocorrelation function for the adiabatic curvature mode and the mode induced by a constant 
comoving magnetic field. 
}
\label{fig1}
\end{figure}

In the numerical solutions background cosmological parameters are chosen to be the bestfit parameters of the base model using Planck  2018 data only \cite{Planck-2018}: $\Omega_bh^2=0.022383$, $\Omega_mh^2=0.14314$, $A_s=2.101\times 10^{-9}$ , $n_s=0.96605$ and $H_0=67.32$ km s$^{-1} $Mpc$^{-1}$.
In figure \ref{fig1}  the equal time power spectrum determining the cross correlation 2-point function (cf. equation (\ref{adL1}))  is shown for different choices of magnetic field parameters at present.
In figure \ref{fig1}  ({\sl upper panel on the left}) it can be seen that the adiabatic and magnetic modes are anticorrelated.
For small wave numbers it is completed dominated by the decaying behaviour of the adiabatic curvature mode until it reaches a minimum and rises again for large wave numbers due to the behaviour of the magnetic mode.
As can be appreciated from figure \ref{fig1}  ({\sl upper panel on the right}) there is a local maximum in the ratio of the absolute value of the cross correlation spectral function over the auto correlation  spectral function of the total matter power spectrum. This is located around the wave number when the  magnetic mode starts dominating over the adiabatic mode matter power spectrum. 
For comparison  relative errors $\frac{\Delta P_m}{P_m}(k)$ of data from the  BOSS DR9 Ly-$\alpha$ forrest \cite{bossLyalpha} and  the SDSS  \cite{sdss} are included. However, these do not provide constraints for  the choice of magnetic field parameters used here.
Since the cross correlation spectral function is negative it lowers the contribution of the constant comoving magnetic field mode. This is best visible 
for magnetic fields with $B_0= 4$nG and $n_B=-2.2$ in the {\it lower panel} of figure \ref{fig1}
in the distinct lines around the wavenumber of equality between the adiabatic and magnetic mode contributions  to  the total matter power spectrum without  ({\it black dashed line})   and with the cross correlation term ({\it magenta dashed-dot-dot  line}).

\begin{figure}[h!]
\centerline{\epsfxsize=4in\epsfbox{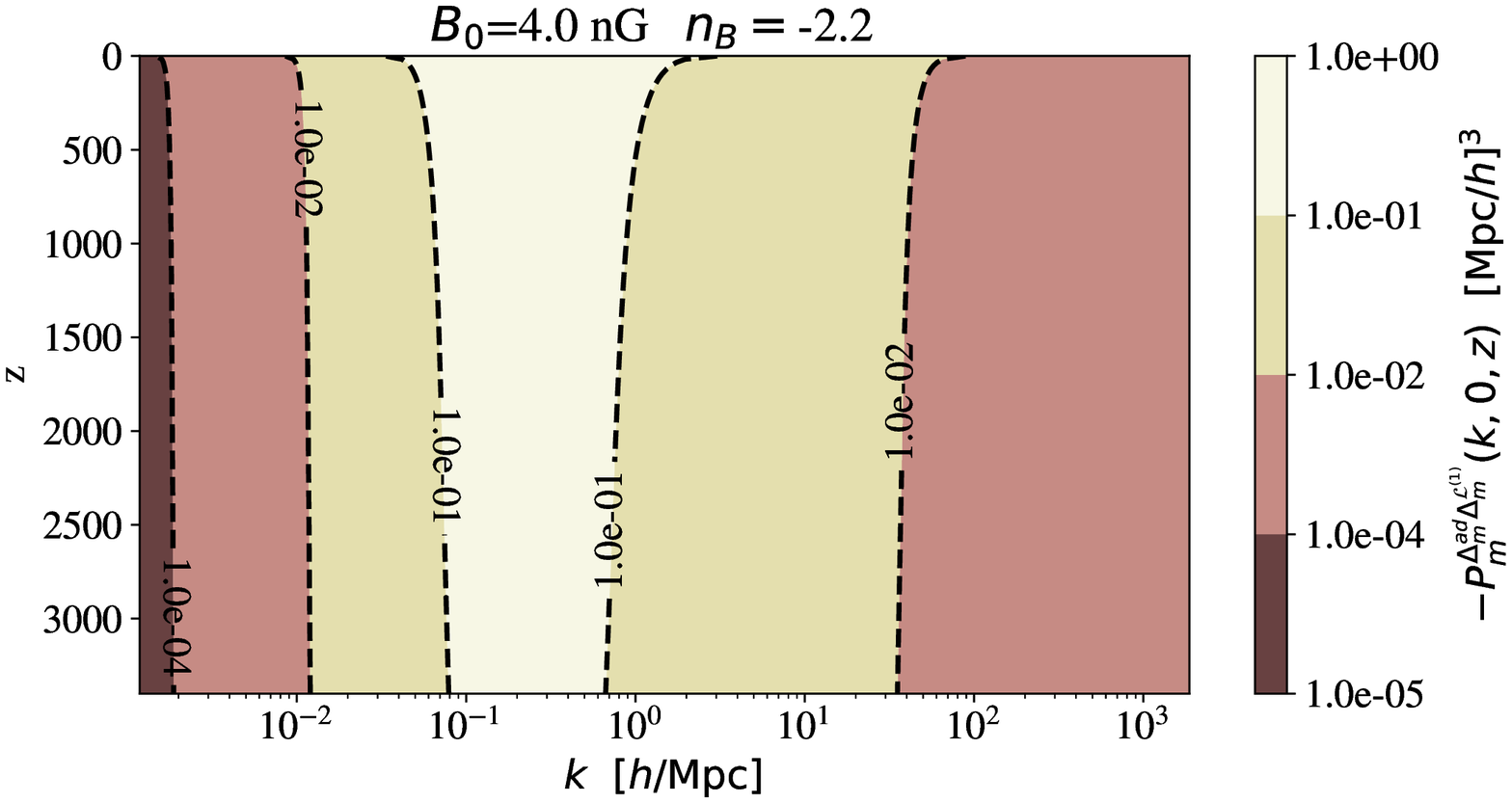}
%\centerline{\epsfxsize=4in\epsfbox{fig-PmadL1_hMpc-1-k_0_z_B(4.0)nB(-2.2).eps}
\hspace{0.1cm}
\epsfxsize=3.4in\epsfbox{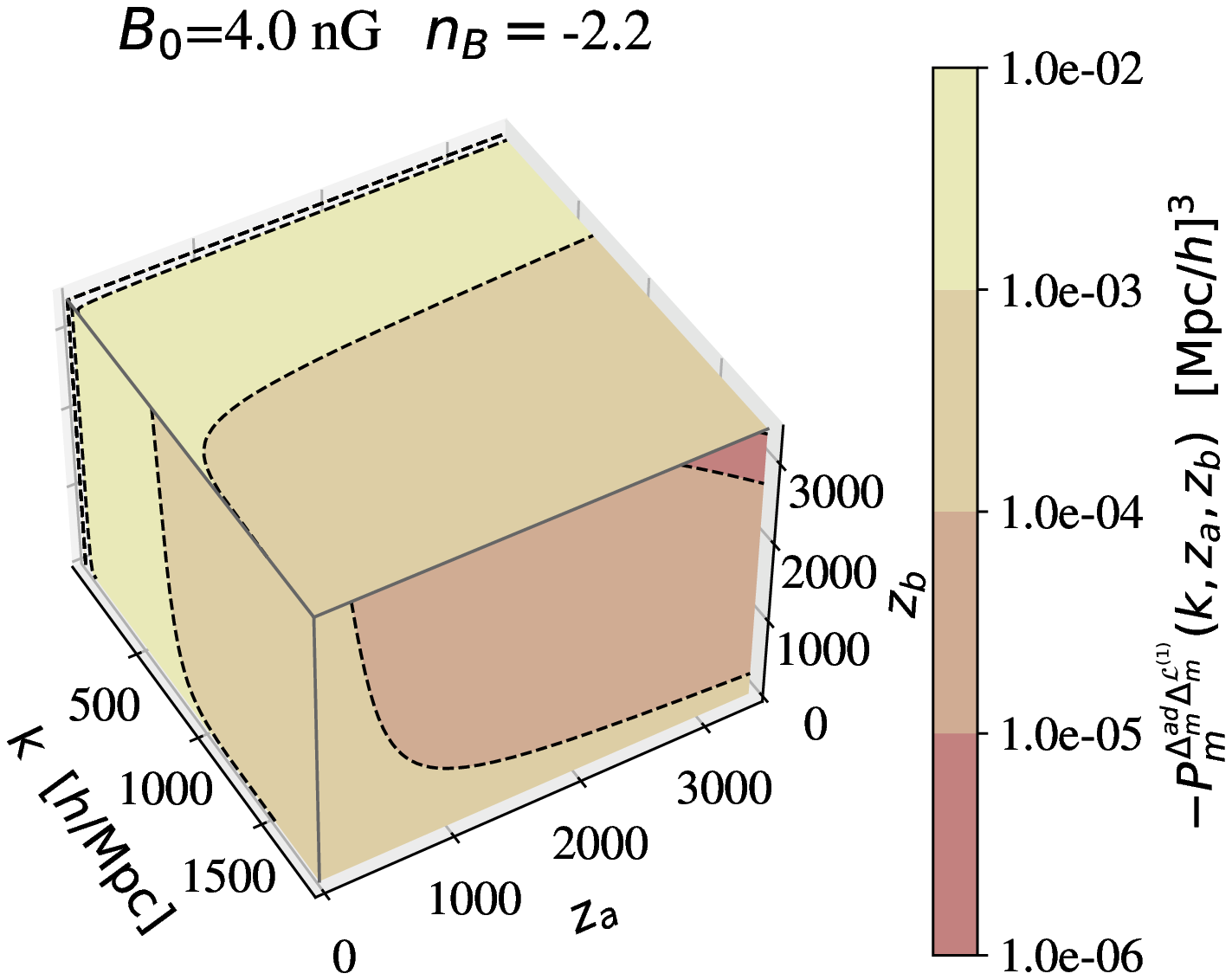}}
\caption{Contour plots of the cross correlation spectral function  of the matter density perturbations due to the adiabatic mode and the evolving magnetic field mode for $B_0=4$ nG and $n_B=-2.2$ at the present epoch, 
$P_m^{\langle\Delta_b^{(ad)}{\cal L}^{(1)}\rangle}(k,z_a,z_b)$.
It is shown for redshifts $z_a=0$ and $z_b=z$  ({\it left}) and the general case for varying redshifts $z_a$ and $z_b$  ({\it right}). 
}
\label{fig2}
\end{figure}

In figure \ref{fig2} the cross correlation function is shown with varying redshift for present day magnetic field strength $B_0=4$ nG and spectral index $n_B=-2.2.$ for which figure \ref{fig1} indicates the largest effect on small scales.
As can be seen in figure \ref{fig2} the cross correlation spectral function is negative across the redshift space spanned by
 $z_a$ and $z_b$. 
 
\begin{figure}[h!]
\centerline{\epsfxsize=3.75in\epsfbox{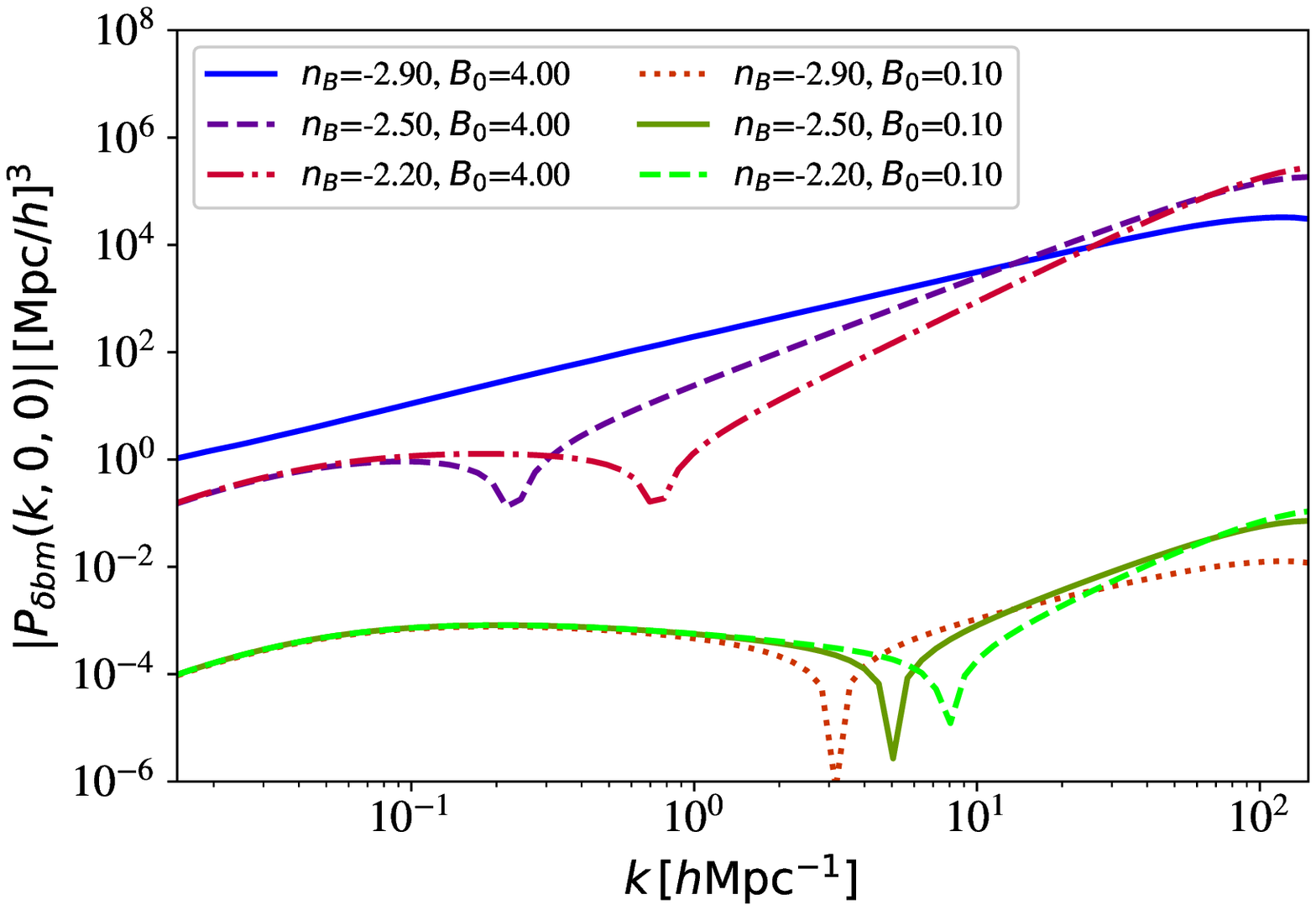}
%\centerline{\epsfxsize=3.75in\epsfbox{1fig-Pdelbm-eqTime_varBnBkhMpc.eps}
\epsfxsize=3.75in\epsfbox{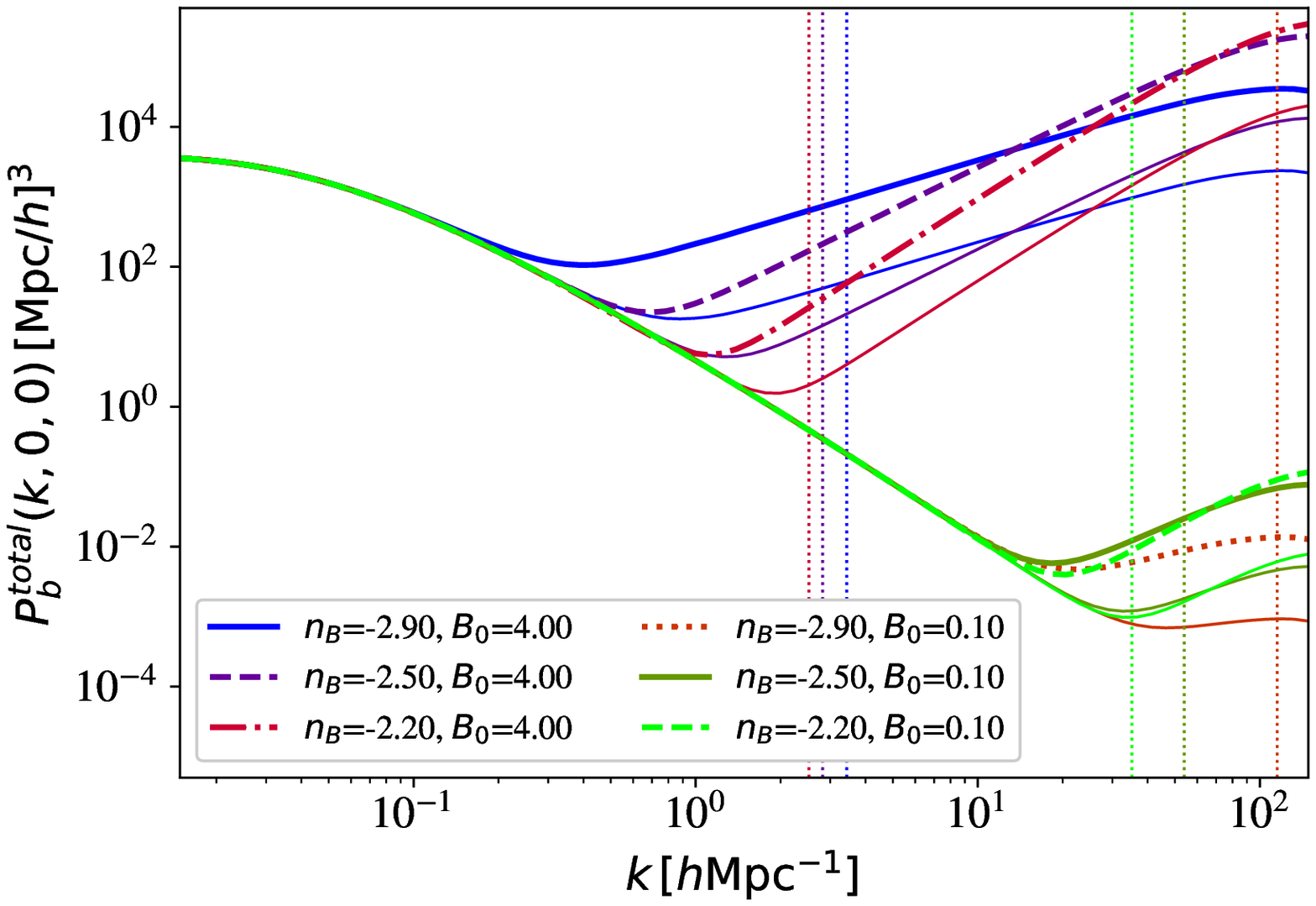}}
\caption{ {\it Left:} The equal time spectral function  $P_{\delta b m}(k,0,0)$ at present for magnetic field parameters ($B_0$ [nG], $n_B$).
{\it Right:} The spectral function determining the present day equal time  auto correlation function of the total baryon density perturbation is shown together with  the corresponding solutions of the total matter power spectral functions (thin solid lines). The dotted vertical lines indicate the positions of the magnetic Jeans wave number for the different choices of magnetic field parameters.}
\label{fig3}
\end{figure}

\begin{figure}[h!]
\centerline{\epsfxsize=4in\epsfbox{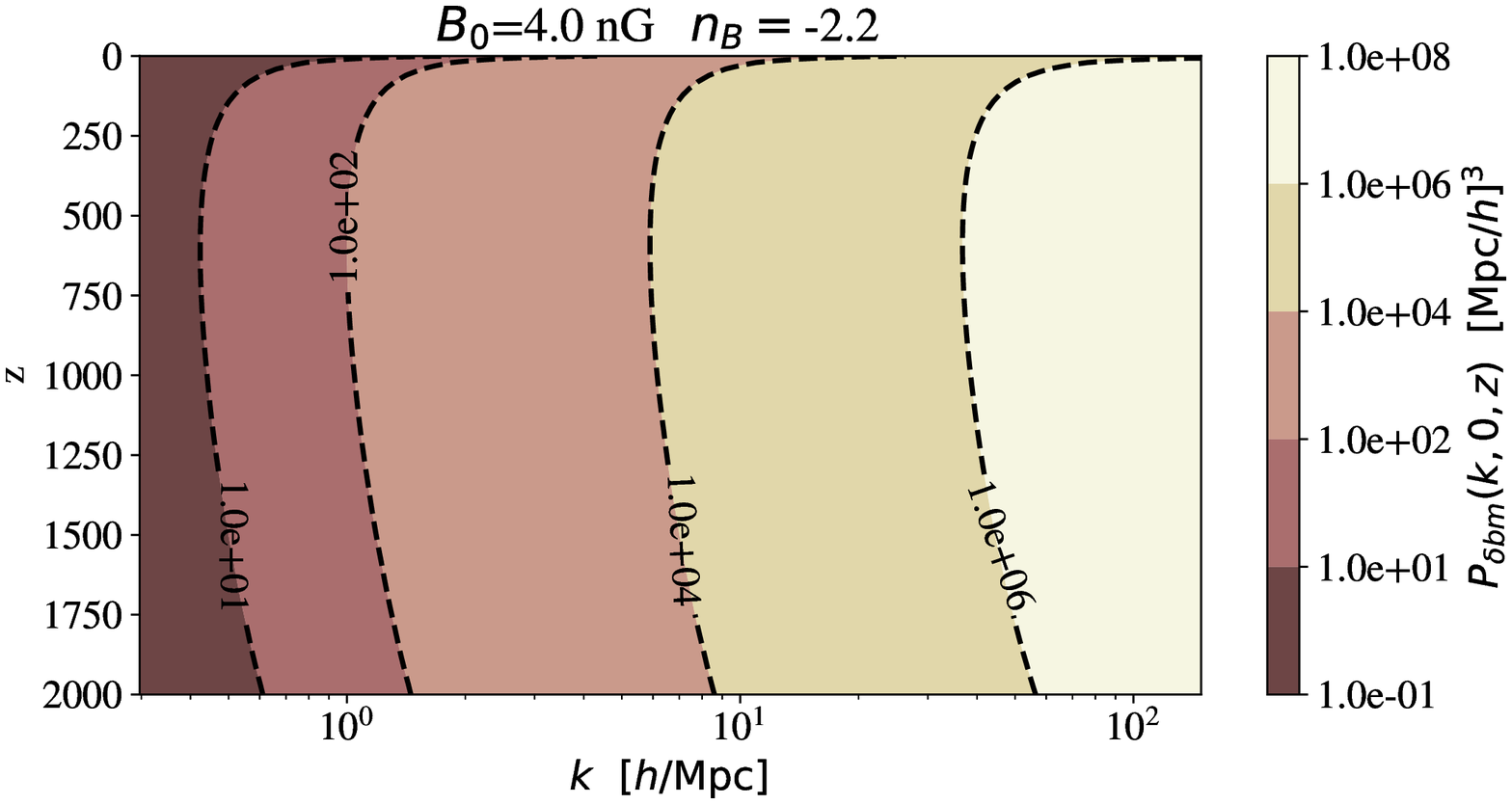}
%\centerline{\epsfxsize=4in\epsfbox{fig-Pdelbm_hMpc-1-k_0_z_B(4.0)nB(-2.2)-new.eps}
\epsfxsize=3.4in\epsfbox{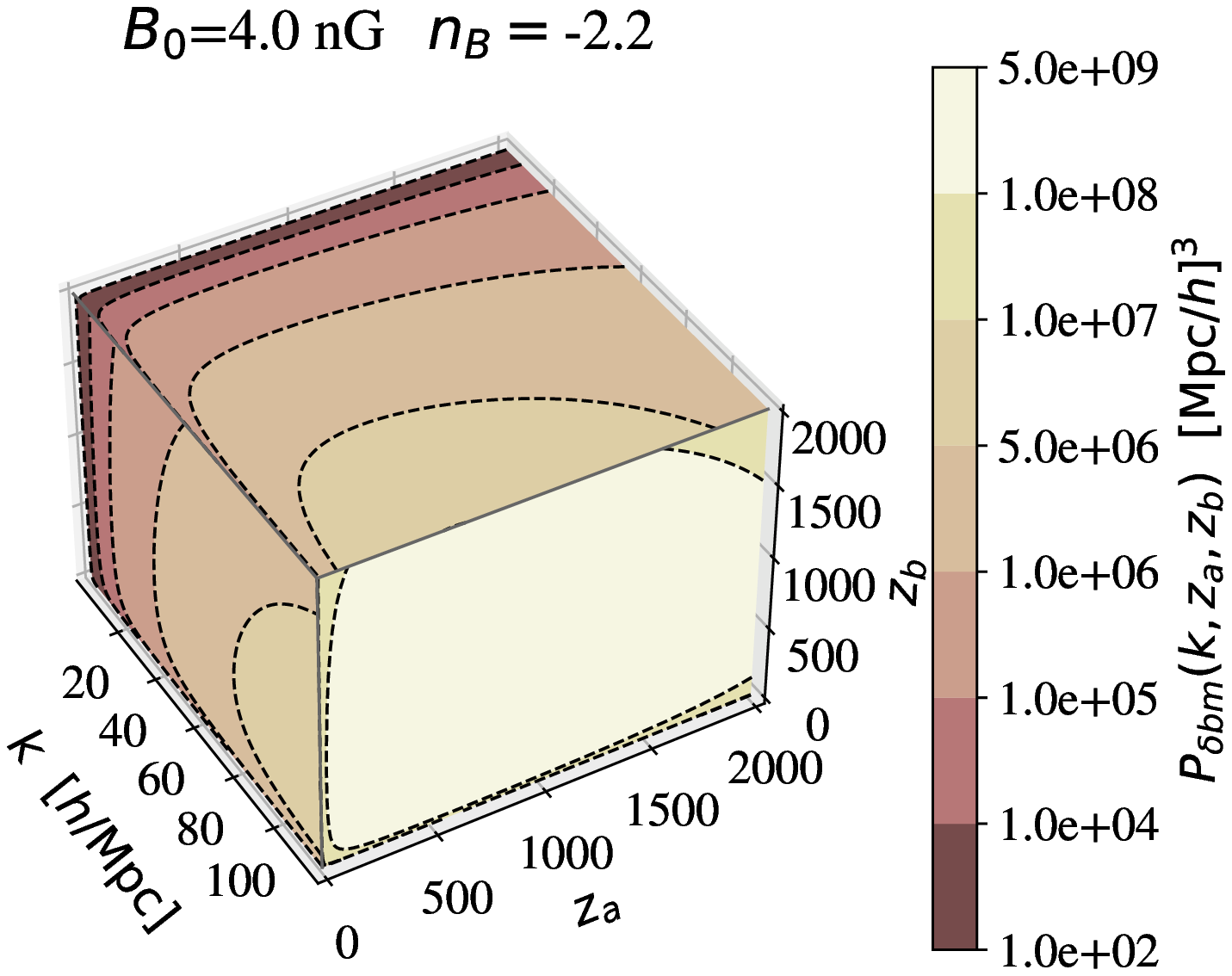}}
\caption{The spectral function $P_{\delta b m}(k,z_a,z_b)$  is shown for the cross correlation between redshift zero and arbitrary redshift ({\it left}) and the general 
case for two arbitrary redshifts  ({\it right}) for $B_0=4$ nG and $n_B=-2.2$.
}
\label{fig4}
\end{figure}

The baryon density perturbation evolves according to (e.g., \cite{kk14})
\begin{eqnarray}
\ddot{\Delta}_b+\frac{2}{\eta}\dot{\Delta}_b=\frac{6}{\eta^2}\Delta_m
-\frac{k^2}{3}\frac{\Omega_{\gamma,0}}{\Omega_{b,0}}\frac{\eta_0^2}{\eta^2}L({\mathbf k},\eta)
\label{Db}
\end{eqnarray}
Solving equation (\ref{Db}) together with equation (\ref{Dmtotal}) shows that the amplitude of the growing mode of the baryon component follow in general an evolution different from the the total matter perturbation. 
The growing mode of the total baryon density perturbation yields  the  auto correlation function  of the total baryonic density perturbation at redshifts $z_a$ and $z_b$.  This can be written as
\begin{eqnarray}
\langle\Delta_b^*({\bf k}',z_a)\Delta_b({\bf k},z_b)\rangle&=&
\langle\Delta_m^*({\bf k}',z_a)\Delta_m({\bf k},z_b)\rangle+
\delta_{{\bf k}, {\bf k}'}D(z_a)D(z_b)P_{\delta b m}(k,z_a,z_b)
\end{eqnarray}
where $D(z)$ denotes the standard growth factor as above. Moreover,
\begin{eqnarray}
P_{\delta b m}(k,z_a,z_b)=
P_m^{(B^{(0)})}(k)\left[
F_1(z_a)+F_1(z_b)+F_1(z_a)F_1(z_b)
\right]
\nonumber\\
+
\left(\frac{\rho_{B,0}}{\rho_{\gamma,0}}\right)\left(\frac{k}{a_0H_0}\right)^2P_m^{(ad)}(k)
{\cal I}_{\langle\Delta_b^{(ad)}{\cal L}^{(1)}\rangle}(k)
\left[F_2(z_a)+F_2(z_b)\right]
\label{delbdelb}
\end{eqnarray}
and
\begin{eqnarray}
F_1(z)&=&
10
\left(\frac{\Omega_{m,0}}{\Omega_{b,0}}-1\right)\frac{1+z}{\sqrt{1+z_{eq}}}
\left[\frac{1}{2}\ln\left(\frac{1+z_{eq}}{1+z}\right)-1+\left(\frac{1+z}{1+z_{eq}}\right)^{\frac{1}{2}}\right].
\nonumber
\\
F_2(z)&=&2\frac{\Omega_{\gamma,0}}{\Omega_{m,0}}
\left(\frac{\Omega_{m,0}}{\Omega_{b,0}}-1\right)
\frac{1+z}{1+z_{eq}}\left[\frac{1}{6}\frac{1+z_{eq}}{1+z}+\frac{1}{2}-\frac{2}{3}
\left(\frac{1+z}{1+z_{eq}}\right)^{\frac{1}{2}}-\frac{1}{2}\ln\left(\frac{1+z_{eq}}{1+z}\right)\right]
\end{eqnarray}
In figure \ref{fig3} the equal time spectral function  $P_{\delta b m}(k,0,0)$ at present as well as the spectral function determining the present day equal time  auto correlation function of the total baryon density perturbation are  shown for different choices of the magnetic field spectral parameters.
In the power spectra of the total baryon density auotcorrelation functions at present (figure \ref{fig3} ({\it right})) the magnetic Jeans  wave number \cite{sesu}
\begin{eqnarray}
\left(\frac{k_J}{{\rm Mpc}^{-1}}\right)=\left[14.8\left(\frac{\Omega_m}{0.3}\right)^{\frac{1}{2}}\left(\frac{h}{0.7}\right)
\left(\frac{B}{10^{-9}{\rm G}}\right)^{-1}\left(\frac{k_L}{{\rm Mpc}^{-1}}\right)^{\frac{n_B+3}{2}}\right]^{\frac{2}{n_B+5}}.
\end{eqnarray}
has been indicated by vertical lines for a magnetic field pivot scale $k_L=1$ Mpc$^{-1}$.
In a pure baryonic matter universe the magnetic Jeans scale provides a limit
beyond which  magnetic pressure prevents further collapse. 
However, since observations show that matter is dominated by cold dark matter rather than baryonic matter the magnetic Jeans scale  does not provide a sharp cut-off in the matter power spectrum \cite{KimOlinRos}.  
In figure \ref{fig4} the spectral function $P_{\delta b m}(k,z_a,z_b)$  which marks the different evolution of the baryon density perturbation from the total matter density perturbation is shown for the cross correlation between redshift zero and arbitrary redshift (figure \ref{fig4} ({\it left})) and the general 
case for two arbitrary redshifts (figure \ref{fig4} ({\it right})).

\section{Conclusions }
\label{s4}
\setcounter{equation}{0}

In general primordial magnetic fields are treated as stochastic magnetic fields frozen into the background fluid
when considering their effect on  the cosmic microwave background or the linear matter power spectrum.
Therefore they only decay with the square of the scale factor as the universe expands. Taking into account the induction equation leads to a non trivial evolution with time of the magnetic field as well as a non vanishing cross correlation between the adiabatic, primordial curvature matter density perturbation and the compensated magnetic mode. The implications of this backreaction on the angular power spectra of the CMB temperature anisotropy and polarization was  already studied in \cite{kk13}.  
Here the effect on the linear matter power spectrum has been considered.  It was found that 
the largest effect on the  matter power spectrum is around the wave number of equality of the adiabatic mode and magnetic mode contributions. For magnetic fields with amplitudes of $B_0=4$ nG and spectral indices $n_B$  with values -2.9, -2.5, -2.2 the cross correlation function reaches appreciable values at the local maximum of the absolute value.
The largest spectral index has the strongest effect.
It is interesting to note that the cross correlation spectral function of the adiabatic and magnetic modes becomes a function redshifts.

\section{Acknowlegements}

Financial support by Spanish Science Ministry grant PGC2018-094626-B-C22 (MCIU/AEI/FEDER, EU) and Basque Government grant IT979-16 is gratefully acknowledged. The author would like to thank the Max-Planck-Institute for Astrophysics for hospitality during the final stages of this work.

%%%%%%%%%%%%%%%%%%%%%%%%%%%%%%%%%%%%%%%%%%%%%%%%%%%%%%%%%%%%%%
%%%%%%%%%%%%%%%%%%%%%%%%%%%%%%%%%%%%%%%%%%%%%%%%%%%%%%%%%%%%%%

\bibliography{references}

\bibliographystyle{apsrev}

\end{document}